\documentstyle[aps,twocolumn,epsfig,rotating]{revtex}

\begin{document}

\draft \preprint{TRIUMF/McMaster U.}
\twocolumn[\hsize\textwidth\columnwidth\hsize\csname    %<--------
@twocolumnfalse\endcsname                               %<--------
\vspace{15mm}
\begin{title} {\bf Population fragmentation and party dynamics in an evolutionary political game}

\end{title}

%\vspace{1truecm}

\author{Arne Soulier and Tim Halpin-Healy}

\address{Physics Department, Barnard College, Columbia University, NY NY 
10027-6598} \vspace{-5truemm}

\vspace{-4truemm}

\date{May 15, 2003}

\maketitle

\begin{abstract}

  We examine kinetic symmetry breaking phenomena in an evolutionary political game in which voters, inhabiting a 
multidimensional
  ideological space,  cast ballots
via selection mechanisms subject to the competing social forces of conformity and dissent. Our understanding of the
spatiotemporally complex population dynamics is informed by a system of nonlinear replicator 
equations, discrete deterministic cousin of the original stochastic Seceder Model.

\end{abstract} 

%\vspace{5mm}

%\pacs{PACS:87.23.Kg, 87.23.Cc, 05.65.+b}
%05.70.Fh Phase transitions: general aspects
%05.50.+q Lattice theory and statistics; Ising problems
%05.50.+q Lattice theory and statistics;

\vskip2pc]                                              %<---------

%%%%%%%%%%%%%%%%%%%%%%%%%%%%%%%%%%%%%%%%%%%%%%%%%%%%%%%%%%%%%%%%
\narrowtext

Fragmentation, dispersal \& coalescence are crucial processes in a multitude of highly correlated, 
dynamically complex many-body systems, particularly those far from equilibrium where
stochasticity, symmetry, and self-organization can conspire to generate rich pattern formation 
phenomena
~\cite{pattern}. Natural examples abound- the rings of Saturn, its Cassini and lesser divisions, consist of  a 
gravitationally 
bound myriad of ever-colliding rocky chucks, macroscopically stable within an intricately detailed set of circular 
bands~\cite{saturn}. Alternatively, consider the coat of the leopard, zebra, or giraffe- each 
a glorious product of competing reaction-diffusion biochemistry ~\cite{murray}. In these instances, the dynamical forces 
are physiochemical in nature, but movement can be inspired by equally compelling, though fundamentally different 
mechanisms, especially in biological, economic and sociological contexts. Visual/verbal cues (i.e., 
information propagation)  within a school of fish~\cite{fish} or flock of birds~\cite{birds} lead to dynamic domain 
formation \& destruction. 
Volatile populations of informed, interacting stock market traders can condense, exhibiting herd-like behavior 
%as well as the splintering of investment strategies
~\cite{herd}. Rich stochastic 
dynamics, as well as phase transition phenomena, are evident in various evolutionary minority (e.g., {\it El Farol} Bar) 
~\cite{zhang}, public-goods ~\cite{pgg}, and other societal selection games, such as the Seceder 
Model~\cite{Seceder,SHH}, which introduces a novel dynamical frustration via competing tendencies of being distinct, yet 
part of the group. Here, we consider the seceder mechanism within an evolving political populace, uncovering an interwoven 
set 
of strange attractors whose reign is governed by kinetic symmetry-breaking within an ideological plane.  
We find that the conflicting tendencies  toward conformity and dissent yield rich spatiotemporal party population 
dynamics.  
The seceder
Catch-22 is that dissenting parties provide alternatives, but frequently grow in popularity themselves, inevitably 
suffering 
rebellion from within.  
Clearly, these 
simple models are not intended to capture
all important features; the goal, however, is to gain appreciation of robust, universal aspects, as was done, e.g.,  
for Ising systems in the case of ferromagnetic, liquid-gas, and binary alloy critical phenomena~\cite{PT}, or
quadratic maps \& Lotka-Volterra coupled ODEs, in the matter of chaotic dynamics~\cite{CD}.

For ease of presentation, we discuss the illustrative case of 3-party 
dynamics driven by a triplet-selection mechanism; i.e., voters poll three individuals to inform their political decision.
We are lead, initially, to study the fixed points (FPs) associated with the following set 
of {\it symmetry-broken} replicator equations ~\cite{Seceder,SHH,HS}:

$\dot R= R^3 +3R(G^2+B^2) + \alpha RGB -R$

$\dot G= G^3 +3G(R^2+B^2) + \beta RGB -G$

$\dot B= B^3 +3B(R^2+G^2) + \gamma RGB -B$

\noindent where $R,G,\& B\le 1$ represent the relative populations of the three parties, Red, Green \& Black,  while 
probability 
conservation demands $\alpha$+$\beta$+$\gamma$=6. The rate variables are 
dictated by the competing tendencies of, resp.,  homogeneity, distinction, and the democratic 
dilemma- equal representation of the three parties.  More specifically, the purely cubic term
results when the triplet selection group has uniform representation of one party; the decision maker
switches to (or remains in, as the case may be...) that party.  In the event that the polled group
has an unequal distribution; i.e., a minority party, outnumbered 2 to 1, the 
underdog triumphs- thus, the seceder linear-quadratic terms.  Finally, the trilinear contributions arise when 
the opinion group has one individual from each of the three parties.  
We consider successive levels of symmetry breaking- first, if $\alpha$=$\beta$, we have effectively, an 
isosceles geometry, $R$ \& $G$ are equivalent in their relation to $B$ and we find the following 
population FPs: $(R,G,B)=({1\over 2}, {1\over 2},0)$ and $({2\over {8-\alpha}},{2\over {8-\alpha}},
{{4-\alpha}\over {8-\alpha}}).$  For $\alpha<4$, the 3-party solution is stable, 2-party dynamics 
unstable; for $\alpha\ge4$ vice versa. Single party domination, e.g., $(0,0,1),$ is never stable.
Clearly, $\alpha=2$ would represent a fully symmetrized situation with the three groups equidistant 
from each other,  
an equilateral (EQ) arrangement with evenly distributed 
populations $({1\over 3},{1\over 3},{1\over 3}).$ By contrast,  $\alpha=3$ corresponds to the standard 
Seceder Model in one dimension ~\cite{Seceder}, with the odd party, $B,$ poised at the midpoint of the line 
segment connecting ideologically opposed groups, $R$ \& $G,$ at the ends; in this case, the 
stable population FP is the familiar $({2\over 5},{2\over 5},{1\over 5}).$ In fact, any short 
isosceles (SI) triangle arrangement of the parties, with sides $RG$$>$$RB$$=$$GB,$ is drawn to a chaotic orbit
about that fixed point; see later. Alternatively, the geometric arrangement could be tall isosceles (TI),
with $RG$$<$$RB$$=$$GB,$ so that there is effectively a remote party equidistant from the others; 
here, connection to the stochastic Seceder Model
suggests $\alpha$=0; the dynamical

\begin{figure}
\begin{center}
  {
  \begin{turn}{0}%
    {\epsfig{file=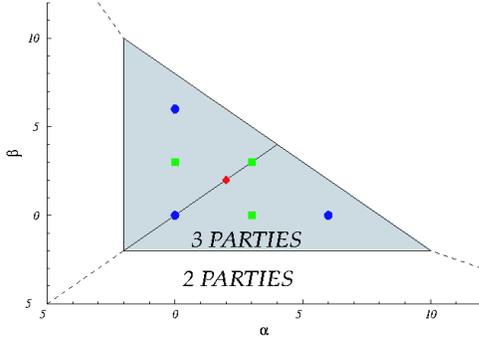,height=4.8cm,width=6.4cm} }
   \end{turn}
   }
\caption{ Phase diagram of the {\it broken-symmetry} replicator equations for 3-party dynamics.}
\end{center}
\end{figure}

\noindent   flows are drawn to the attractor $({1\over 4},{1\over 2},{1\over 
4}).$   
We reveal, in Figure 1, the full phase-diagram of this nonlinear replicator system in 
the $\alpha\beta$-plane, relaxing the isoscelean geometric constraint.   Surprisingly, a stable 3-party dynamic 
persists for quite a range of parameter
values, occupying the triangular region bounded by the lines $\alpha=-2, \beta=-2, \alpha + \beta=8$ (i.e., 
$\gamma=-2$). The unique, full symmetry EQ FP, 
$({1\over 3},{1\over 3},{1\over 3}),$ shown as a red diamond at the point $\alpha=\beta=\gamma=2,$ is surrounded 
by a trio of permutation-related TI (SI) FPs, indicated by blue circles (green squares). Oddly enough, 3-party politics 
exists even for modestly negative values 
of $\alpha, \beta,$ or $\gamma$; i.e., explicit biasing against a particular party. For sufficiently great forcing, 
however, 
the dynamics switches from 3 to 2-party politics; e.g., with $\gamma<-2,$ all flows are
driven to the same $(R,G,B)=({1\over 2}, {1\over 2},0)$ FP, with a 50-50 split in the population 
between the two surviving parties.  Along special lines of symmetry, such as $\alpha=\beta<-2,$
there is, numerically, a first-order coexistence between competing relevant 2-party FPs; in the 
mentioned instance, $(0,{1\over 2}, {1\over 2})$ and $({1\over 2},0,{1\over 2})$.  
All this is in strong contrast to the 3-party region of the phase diagram, where there is a
unique FP associated with each choice of $\alpha\beta\gamma$ values.

Having investigated the role of symmetry-breaking within the context of the {\it well-mixed, deterministic}
replicator equations, we now explore its manifestation for a related {\it spatially-extended, stochastic model} in which, 
one might imagine, voters occupy specific positions in a 2D ideological plane, with axes corresponding to a pair of 
compelling political issues,
e.g., taxation and international cooperation, running the gamut from extreme leftist to ultra-right wing conservative 
positions.
We'll see shortly that for this stochastic Seceder Model, kinetic symmetry breaking (KSB) arises 
in an entirely intrinsic manner via  
fluctuations built-up in this far-from-equilibrium system. 
We make long-time runs, $t$=5x10$^5$ generations, with substantial population sizes, $N$=10$^3,$
running simulations in 2d with the entire population initialized in a single party at the origin. 
As discovered previously ~\cite{SHH},  the population rapidly fragments and disperses but within a 
thousand generations coalesces into three divergent groups, heading radially away from 
the origin 
at a slightly {\it sublinear} speed~\cite{natalie} with  equally-sized

\begin{figure}
\begin{center}
  {
  \begin{turn}{0}%
    {\epsfig{file=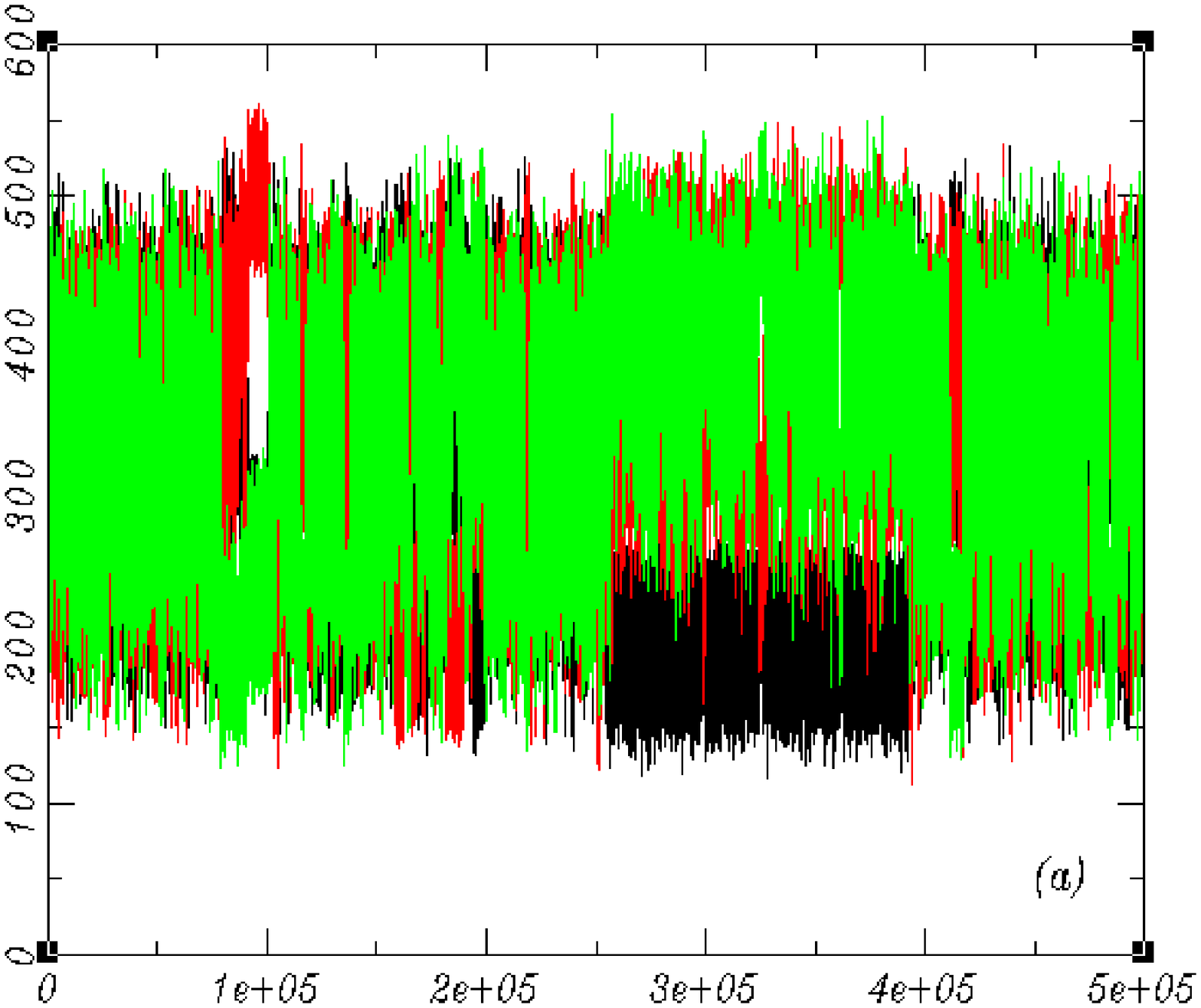,height=6.3cm,width=8.4cm} }
   \end{turn}
   }
   {
  \begin{turn}{0}%
    {\epsfig{file=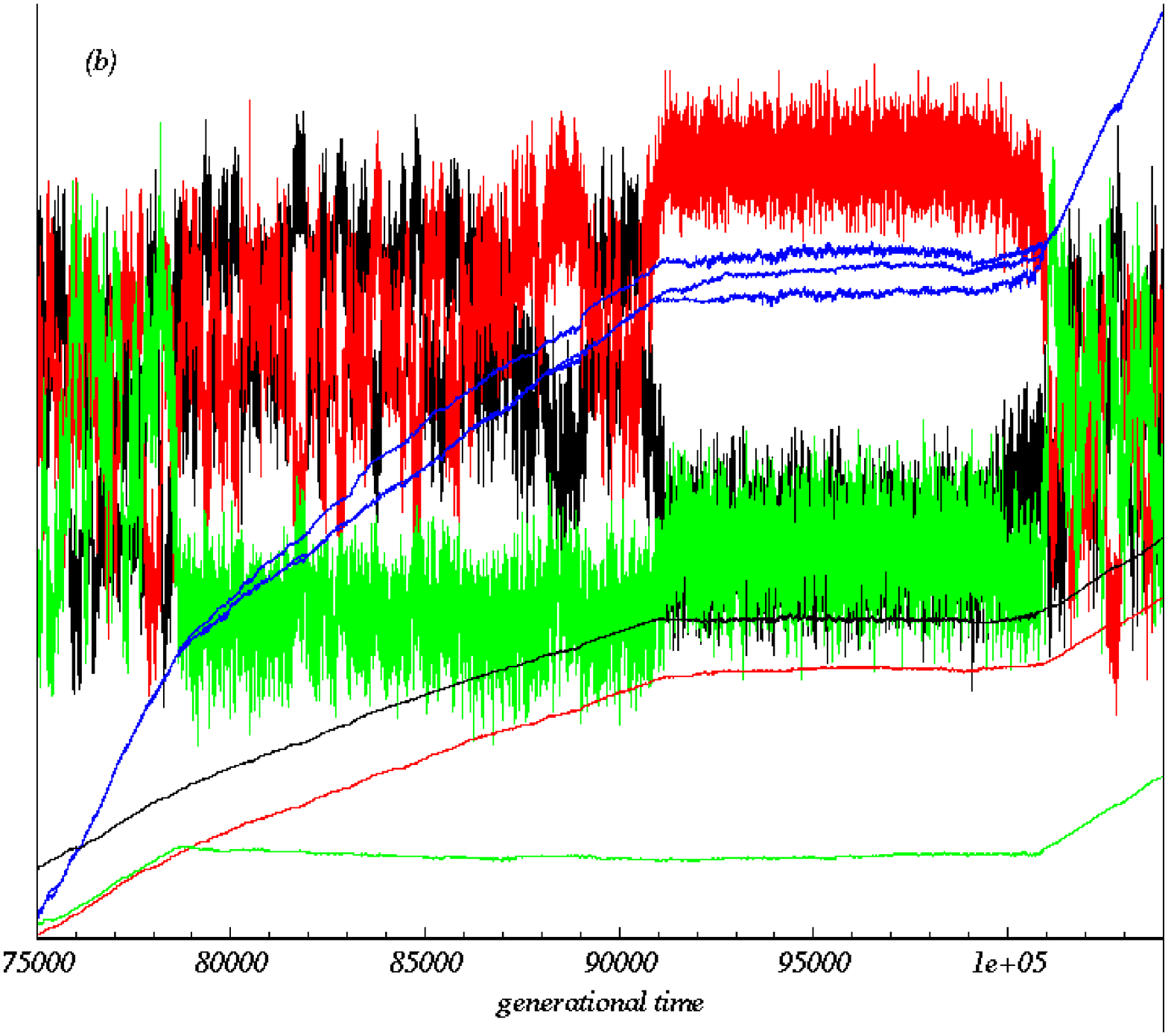,height=5.4cm,width=7.2cm} }
   \end{turn}
   }
\caption{a) Green, Red, \& Black Party membership fluctuations over a half million 
generations, 
and b) magnified view of a particularly 
busy 30000 generation time interval, $t=75-105$x$10^3,$ when the population circulates about EQ, SI, TI 
FPs. Also shown are the pairwise separations (blue) and 
party radial distances, in the latter instance w/ 
eponymous traces.}
\end{center}
\end{figure}

\noindent   party 
populations. In fact, if one carefully monitors party membership over time, there 
are impressive fluctuations
about (as well as away from!) the neighborhood of this dynamical $({1\over 3},{1\over 3},{1\over 3})$  EQ FP.  In Figure 
2a, we 
show the entire time series for the three parties- here indicated by the colors red, green 
and black.  We see immediately 
that for much of $t<1$x10$^5,$ the average party population is indeed $N/3$ but in due 
course, one of the parties, in this case the Greens @$t$$\approx$78000, suffers a severe drop in membership, suddenly 
falling to 
200; simultaneously, the Red and Black Parties rise, fluctuating about a mean of 400. In Figure 2b, we show an expanded 
detail 
of the subsequent, especially active epoch. Apparently, the political system has dynamically migrated to a chaotic stay 
about the $({2\over 5},{1\over 
5},{2\over 5})$ SI FP, where two parties, Red \& Black, equally dominate the third, Green. This state of affairs persists 
until $t$$\approx$91000, when a second abrupt dynamical transition occurs- this time to the $({1\over 2},{1\over 
4},{1\over 4})$ TI FP, in which there is 
a single popular party, here the Reds, with twice the membership of two equal minority parties, 
Black and Green.  Finally, just beyond $t$=100000, the system reverts to the original, unique EQ FP. Overlays to Figure 
2b reveal relevant geometric
aspects, including evolution of various pairwise separations 
between parties (blue), as well as radial distances from the origin (party colors). We notice, for example, when the 
Green Party membership plummets,  the party actually suffers a {\it stochastic stall} (green trace plateaus...), its 
radial motion ceasing as population redistribution takes place.  The SI$\rightarrow$TI transition is always marked by the 
simultaneous stall of all three parties (red, green, and black traces all horizontal), while the triangular geometry 
switches to one short and two (nearly) equal long sides.  The return the EQ FP for $t$$>$100000 is, of course, temporary-
see, again, Figure 2a. Note, in particular, the rather long run, $t$=2.5-3.9x10$^5$, during which the political dynamic 
is controlled by the $({2\over 5},{2\over 
5},{1\over 5})$ SI FP, though even that impressive interlude is punctuated by  intermittent, though typically short-lived 
($\approx$ 1-2000 generations) TI episodes, three of which are clearly visible in the figure.

It is informative to study the time evolution of the party populations from the vantage 
point of a Poincare type section; we show in Figure 3 the fractional 
populations of the Black and Red Parties for the entirety of the original simulation.
For the time interval $t$=2.0-4.0x10$^5,$ an epoch strongly 
dominated, recall Figure 2a, by the $(R,G,B)=({2\over 5},{2\over 5},{1\over 5})$ FP, indicated by leftmost green square, 
a symbol shared by its SI FP permuted cousins. As previously, the TI FPs 
are shown by blue circles, and the solitary, but evenly distributed, EQ FP is the red diamond. The 
time interval was chosen specifically to show governing role of a particular SI FP, which gives rise 
to a compelling {\it crescent-shaped} feature in the Poincare trace. Note, especially, the 
position of complementary blue TI FPs in the wings of the crescent, which faithfully 
track the relatively less frequent excursions of the system to states in which a single majority 
party strongly dominates the political dynamics; here, Red or Green Parties with memberships in the 
neighborhood of 500- in such instances, the Black Party moves up momentarily to 250. It is apparent 
that the interval  $t$=2.0-2.4x10$^5$ is responsible for the faint black cloud centered about the EQ FP, complementary to 
the solid crescent. For the given interval, the system spends little time 
with either the Green or Red Parties in the weak minority,  so those sections of the Poincare plot are 
rather sparse, but in the infinite time limit, or even if one considers the full 1/2 million 
generations implicit in Figure 2a, the Poincare section (black and gray dots taken together) has an easily perceived 
elliptical form
resulting from the superposition of three symmetrically permuted cusps pinned at each of the green square, SI FPs.  
In this way, wings overlap, where the crescents share a common, blue circle TI FP.
Of course, this ellipse (or ellipsoid, in the full 3d plot with (R,G,B) on the three axes...), does 
not have a sharp boundary, though the fall-off is rather rapid. The soft asymptotic ellipse, as well 
as the short-

\begin{figure}
\begin{center}
  {
  \begin{turn}{0}%
    {\epsfig{file=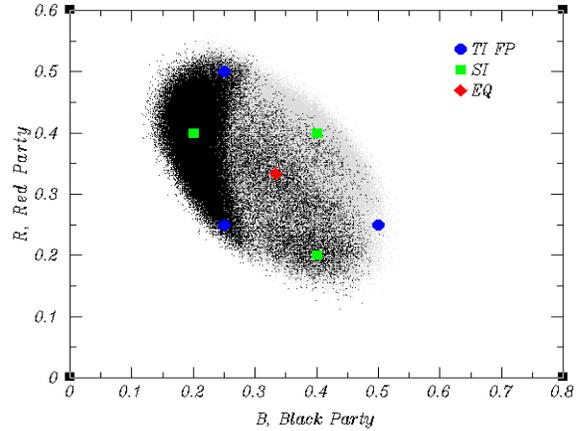,height=6.0cm,width=8.0cm} }
   \end{turn}
   }
\caption{Poincare section tracks the system through successive generations, plotting the population 
fractions claimed by two of the three parties. The black dots correspond to the time interval 
$t$=2.0-4.0x10$^5$ during most of which the Reds and Greens share power, with Black Party in solitary 
minority and, aside from brief intermittent excursions elsewhere, the dynamics are controlled by the SI FP
$({2\over 5},{2\over 5},{1\over 5})$- see Figure 2a. Note the black {\it crescent}-shaped region, with tips overlapping 
neighboring TI FPs in blue. The gray dots record the system in the complementary time interval of the half million 
generation simulation. 
Together, black and grey yield a very particular elliptical geometry, which along with the crescent, are the 
characteristic features 
Seceder Model Poincare section.}
\end{center}
\end{figure}

\noindent timescale crescent constitute the signature details of the Seceder Model Poincare section.

We can decipher these matters a bit by studying the 
party membership PDFs, obtained by isolating intervals of time in which a single FP
dominates the dynamics. In Figure 4,  the population PDFs associated 
with the EQ, SI, TI attractors are shown, resp., in red, green, and blue. 
We construct the red $({1\over 3},{1\over 
3},{1\over 3})$ EQ PDF by sampling over some 200000 noncontiguous generations, primarily $t$$<$80000 and 
$t$$>$420000 and find three overlapped, essentially identical, asymmetric, nongaussian, skewed  distributions, peaked 
near $N/3$=333, low slung with very substantial widths. By contrast, focussing on the 150000 generation interval 
commencing at $t$=240000, where $({2\over 5},{2\over 5},{1\over 5})$ FP reigns, we uncover the 
membership PDF associated  with the green SI FP.  Clearly a different beast altogether,  it is bimodal, with 2 
superposed, short, but 
broad components  (mean@400), 
long tails on the low side- still substantial even at 250, and a single sharp, nearly symmetric peak centered at 200, 
considerably narrower, with rather small probability to rise above 
250 or fall below 150. These features, esp. the latter, are evident in the crescent  boundaries of Figure 3. Finally, the 
TI PDF, in blue, also bimodal, is characterized by a very narrow peak for the solitary 
majority party 
@500, 

\begin{figure}
\begin{center}
  {
  \begin{turn}{0}%
    {\epsfig{file=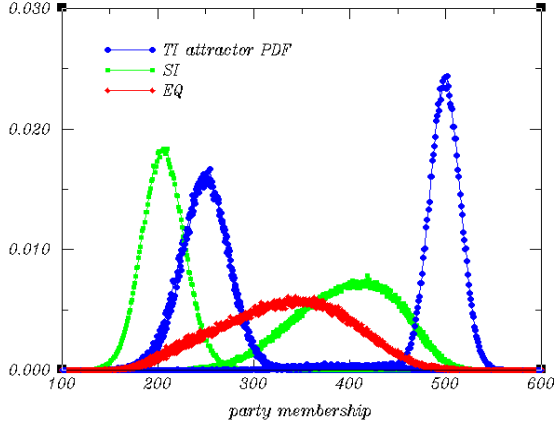,height=6.0cm,width=8.0cm} }
   \end{turn}
   }
\caption{Seceder SI, EQ, TI FP party membership PDFs.}
\end{center}
\end{figure}

\noindent and two identical overlapped broader contributions with mean at 250 for the 
equal pair of minority parties.  The nature of the various tails, as well as the mixed population size, go far in 
explaining why certain dynamical transitions, such as EQ$\rightarrow$SI or SI$\rightarrow$TI, are considerably more 
common than others; e.g., the system
rarely tunnels EQ$\rightarrow$TI. We seek a deeper understanding of the transition probabilities associated with these 
important
global shifts in the population dynamics and have taken a step in this direction, studying the decay modes and 
lifetime distribution of an initial EQ
state- see Figure 5.  We discover that the branching ratio is highly biased in favor of the EQ$\rightarrow$SI decay route; 
indeed,
97.6\%(2.4\%) of the time the transition is to the SI(TI) attractor.     

Interestingly, the essential aspects of KSB can be analyzed via the replicator equations; in 
particular, we consider the case of $d+1$ symmetrically arrayed parties in $d$ dimensional space
(i.e, 2D equilateral triangle, 3D tetrahedron, etc.) and examine what happens if a single party, 
with fractional membership $z$, let's say, kinetically breaks the symmetry, moving closer/farther 
from the remaining parties, each of whom possess equal membership $x=(1-z)/d.$ For this 
characteristic situation, the rate equation for the anomalous party reads: 
$\dot z= z^3 +3dz(1+\delta(d-1))x^2 -z,$ where the symmetry-breaking parameter 
$\delta=0,{1\over 2},1,$ in SI, EQ, TI geometries, with 
associated fixed point values $z^*={{3-d}\over{3+d}},{1\over{d+1}}, {1\over 2}$. The $\delta=
{1\over 2}$ result is expected, the total population distributed evenly amongst the 
$d+1$ groups. 
If, however, the anomalous party distances itself from the rest moving farther away, this 
subpopulation grows to a dimension-{\it independent} percentage of 50\%, the remaining parties sharing 
equally the other half of the population; of course, for the 2D case we've focussed on thus far, 
this corresponds to the TI FP $({1\over 4},{1\over 4},{1\over 2}).$  Although fluctuations can permit 
higher fractions for this dominant party- our earlier simulations indicated, see Figures 2\&4, upwards 
of
3/5, there are apparently dynamical constraints that strongly limit the maximum membership 
of this favored party.  Should the party in power become too fashionable,

\begin{figure}
\begin{center}
  {
  \begin{turn}{0}%
    {\epsfig{file=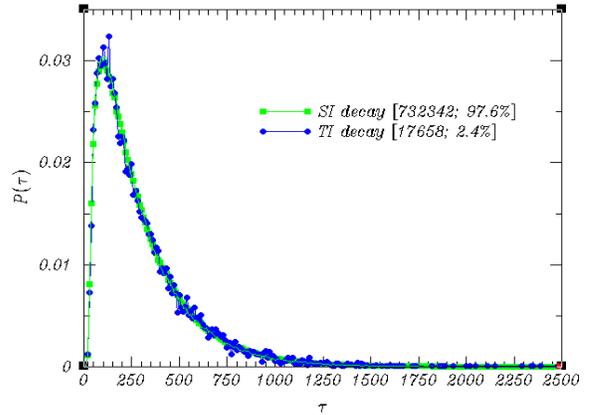,height=6.0cm,width=8.0cm} }
   \end{turn}
   }
\caption{EQ FP lifetime distributions via decay modes to SI and TI dynamical states; 750000 realizations.}
\end{center}
\end{figure}

\noindent the seceder mechanism 
inceases the likelihood of desertion from within. Finally, in the situation where the stochasticity 
draws the errant party closer to the others, a stable SI FP exists only stable below the upper critical 
dimension $d_c=3!$ Consequently, fluctuations in $d\ge3$ bring about the successive demise of all 
but three of the remaining parties, the system suffering cascadal collapse \& dimensional reduction to a 2D ideological hyperplane~\cite{SHH}.

%In Figure, we plot the growth rate $\dot z$ versus fractional party membership 
%for this case of SI KSB. It is clear that for our old friend $d=2,$ the fixed point at $z*={1\over 
%5}$ is stable, but for higher integer dimensions, any flow that begins at finite $z$  flows to the 
%origin, meaning that the membership of this accommodating party diminishes continuously until its 
%ranks are entirely depleted.  

%%%%%%%%%%%%%%%%%%%%%%%%%%%%%%%%%%%%%%%%%%%%%%%%%%%%%%%%%%%%%%%%%%%%%%%%% 
Financial support for Tim HH has 
been provided by NSF DMR-0083204, Condensed Matter Theory.


\vspace{-7mm}
\begin{thebibliography}{99}
\vspace{-14mm}
\bibitem{pattern} for an inspired, if brutally idiosyncratic, entirely algorithmic perspective, see S. Wolfram, {\it 
A New Kind of Science} (Wolfram Media, Champaign IL 2002).
\bibitem{saturn}  J. Schmidt {\it et al.,} Phys. Rev. Lett. {\bf 90}, 061102 (2003).
\bibitem{murray} J. D. Murray, {\it Mathematical Biology} (Springer-Verlag, Berlin 1993).
\bibitem{fish} J. Parrish {\it et al.,} Science {\bf
284}, 99 (1999).
\bibitem{birds} J. Toner and Y. Tu, Phys. Rev. Lett. {\bf 75}, 4326
(1995);  E.V. Albano, 
{\it ibid,} {\bf 77}, 2129
(1996).
\bibitem{herd} V. Egui'luz {\it et al.,} Phys. Rev.
Lett. {\bf 85}, 5659 (2000).
\bibitem{zhang} D. Challet and Y.-C. Zhang, Physica A {\bf 
246}, 407 (1997); N.F. Johnson {\it et al.,} Phys. Rev. Lett. {\bf 82}, 3360 (1999);  S. Hod and E. 
Nakar,
{\it ibid,} {\bf 88}, 238702 (2002).
\bibitem{pgg} G. Szab\'o {\it et al.,} Phys. Rev. Lett. {\bf 89}, 118101 (2002); C. Hauert {\it et al.,} Science {\bf 
296}, 1129 (2002).
\bibitem{Seceder} P. Dittrich, {\it et al.,}   Phys. Rev. Lett. {\bf 84}, 3205 (2000).
\bibitem{SHH} A. Soulier and T. Halpin-Healy, cond-mat/0209451, Phys. Rev. Lett. {\bf 90}, in press.
\bibitem{PT} K. Huang, {\it Statistical Mechanics,} (J. Wiley, NY 1987); C. Domb, {\it The Critical Point,}
(Taylor \& Francis, NY 1996).
\bibitem{CD} J. M. T. Thompson and H. B. Stewart, {\it Nonlinear Dynamics and Chaos,} (J. Wiley, NY 2002). 
\bibitem{HS} J. Hofbauer and K. Sigmund, {\it Evolutionary Games \& Population Dynamics}
	(Cambridge Press, 1998).
\bibitem{natalie} A. Soulier, N. Arkus, and T. Halpin-Healy, submitted.
\end{thebibliography}
\end{document}